\documentclass[journal]{IEEEtran}

\usepackage{cite}
\usepackage{amsmath,amssymb,amsfonts}
\usepackage{algorithmic}
\usepackage{graphicx}
\usepackage{textcomp}
\usepackage{array}

\usepackage[dvipsnames]{xcolor}
\usepackage{multicol}
\usepackage{amssymb}
\usepackage{subfig}
\usepackage{multirow,tabularx,booktabs}
\newcolumntype{P}[1]{>{\centering\arraybackslash}p{#1}}
\DeclareMathOperator{\spn}{\rm{col}}
\usepackage{epstopdf}

\begin{document}

\title{Multi-subject task-related fMRI data processing via a two-stage generalized canonical correlation analysis}

\author{Paris~A.~Karakasis,~\IEEEmembership{Student Member,~IEEE},
        Athanasios~P.~Liavas,~\IEEEmembership{Member,~IEEE},\\
        Nicholas~D.~Sidiropoulos,~\IEEEmembership{Fellow,~IEEE},
        Panagiotis~G.~Simos, and Efrosini Papadaki
    \thanks{P. A. Karakasis was with the School of Electrical and Computer Engineering, Technical University of Crete, Crete, Greece. He is now with the Department of Electrical and Computer Engineering, University of Virginia, Charlottesville, VA 22904,
{\tt\small karakasis@virginia.edu.}}
 \thanks{A. P. Liavas is with the School of Electrical and Computer Engineering, Technical University of Crete, Crete, Greece,
{\tt\small { liavas@telecom.tuc.gr.}}}
\thanks{N. D. Sidiropoulos is with the Department of Electrical and Computer Engineering, University of Virginia, Charlottesville, VA 22904,
{\tt\small nikos@virginia.edu. }}
\thanks{P. G. Simos and E. Papadaki are with the School of Medicine, University of Crete, Crete, Greece and the Institute of Computer Science, Foundation of Research and Technology-Hellas, Crete, Greece,
{\tt\small akis.simos@gmail.com, fpapada@otenet.gr.}}    
\thanks{ P. A. Karakasis, A. P. Liavas, P. G. Simos, and E. Papadaki were partially supported by the European Regional Development Fund of the European Union and Greek national funds through the Operational Program Competitiveness, Entrepreneurship, and Innovation, under the call RESEARCH - CREATE - INNOVATE (project code: T1E$\Delta$K-03360).}
\thanks{ N. D. Sidiropoulos was supported in part by NSF ECCS-1807660, and NSF IIS-1704074.}}


\IEEEoverridecommandlockouts
\IEEEpubid{\makebox[\columnwidth]{978-1-5386-5541-2/18/\$31.00~\copyright2018 IEEE \hfill} \hspace{\columnsep}\makebox[\columnwidth]{ }}

\maketitle
\IEEEpubidadjcol

\begin{abstract}
Functional magnetic resonance imaging (fMRI) is one of the most popular methods for studying the human brain. Task-related fMRI data processing aims to determine which brain areas are activated when a specific task is performed and  is usually based on the Blood Oxygen Level Dependent (BOLD) signal. The background BOLD signal also reflects systematic fluctuations in regional brain activity which are attributed to the existence of resting-state brain networks. 
We propose a new fMRI data generating model which takes into consideration the existence of common task-related and resting-state components. We first estimate the common task-related temporal component, via two successive stages of generalized canonical correlation analysis and, then, we estimate the common task-related spatial component, leading to a task-related activation map. The experimental tests of our method with synthetic data reveal that we are able to obtain very accurate temporal and spatial estimates even at very low Signal to Noise Ratio (SNR), which is usually the case in fMRI data processing. The tests with real-world fMRI data show significant advantages over standard procedures based on General Linear Models (GLMs). 
\end{abstract}


\IEEEpeerreviewmaketitle

\section{Introduction}
\label{sec:introduction}

\IEEEPARstart{F}{unctional} magnetic resonance imaging (fMRI) is a popular approach for studying the human brain. It provides a non-invasive way to measure brain activity, by detecting local changes of blood oxygen level dependent (BOLD) signal in the brain, over time. The aim of task-related fMRI data analysis is to determine which brain areas are activated when a specific task is performed and is usually based on the BOLD signal. Hence, we can create brain activation maps related to specific tasks, which is very useful for understanding the functioning of the human brain. 

In task-related studies \cite{fox2007spontaneous,fair2007method}, it has been noted the presence of spontaneous modulation of the BOLD signal, which cannot be attributed to the experimental excitation or any other explicit input or output and is usually considered as ``noise.''
However, in addition to physiological and magnetic noise, background BOLD signal 
reflects systematic fluctuations in regional brain activity. In particular, BOLD fluctuations are 
correlated between functionally related brain regions, forming resting-state brain networks.
This baseline activity continues during task performance, showing a similar neuro-anatomical
distribution to that observed at rest \cite{fox2006coherent,greicius2004default,arfanakis2000combining,fransson2006default}.
In \cite{fox2006coherent}, it has been suggested that  measured neuronal responses
consist of an approximately linear superposition of task-evoked
neuronal activity and ongoing spontaneous activity.

Various unsupervised multivariate statistical method have been used in 
fMRI data processing. Their aim is to provide information about functional brain connectivity, by describing brain responses in terms of spatial and temporal activation patterns, under no prior assumptions about their form \cite{lindquist2008statistical}. Common multivariate methods which have been used in fMRI data processing are Principal Component Analysis (PCA) \cite{andersen1999principal,hansen1999generalizable}, Independent Component Analysis (ICA) \cite{mckeown1998independent,calhoun2001spatial,calhoun2003ica,erhardt2011comparison}, and tensor factorization based models
\cite{andersen2004structure,beckmann2005tensorial,stegeman2007comparing,morup2008shift,morup2011modeling,chatzichristos2017higher,chatzichristos2017parafac2,madsen2017quantifying}.

Canonical correlation analysis (CCA) is a well known statistical method \cite{hotelling1936relations}
which can be considered as a generalization of the PCA. It takes as input two random vectors and computes two basis vectors such that the correlation between the respective projections of the random vectors onto the basis vectors is maximized \cite{hardoon2004canonical}. After considering the subspace spanned from this set of basis vectors, CCA can be also considered as a method for the estimation of a linear subspace which is ``common'' to the two sets of random variables \cite{ibrahim2019cell}.

Generalized CCA (gCCA) for more than two random vectors dates back to \cite{vinograde1950canonical,steel1951minimum,horst1961generalized,horst1961relations,carroll1968generalization}. In \cite{kettenring1971canonical}, five different formulations of gCCA are presented with
all of them boiling down to the classical CCA when the number of random vectors is equal to two \cite{asendorf2015informative}. Among the different formulations of gCCA appearing in \cite{kettenring1971canonical}, only the solutions of MAX-VAR and MIN-VAR can be obtained directly via eigen-decomposition; the other formulations lead to nonconvex optimization problems, thus, their reliability is questionable. 

Given a set of $M$ random vectors, the initial goal of gCCA was to extract several $M$-dimensional random vectors, known as canonical variates, consisting of unit variance linear compounds (combinations) of the $M$ random vectors \cite{kettenring1971canonical}. MAX-VAR assumes that each linear compound participating in the same canonical variate is a potentially noisy and scaled version of the same common random variable, while different canonical variates are associated with uncorrelated common random variables. On the other hand, MIN-VAR assumes that each linear compound participating in the same canonical variate is a potentially noisy version of a linear compound of the same common random vector of $M-1$ elements. In this setting, linear compounds of different canonical variates must be uncorrelated. MAX-VAR is easier to interpret, thus it has recently become very popular. Moreover, the simple and elegant MAX-VAR formulation presented in \cite{carroll1968generalization} is equivalent to one described above and offers a solution via eigen-decomposition. This is the formulation we adopt in this work.

\subsection{Problem Definition}

We consider the problem of multi-subject task-related fMRI data processing with only one type of stimulus. 
Our aim is to accurately determine, in a data-driven manner, which brain areas are activated when the stimulus is applied and construct the associated brain activation map. 

\subsection{Related Work}

In \cite{li2008cca}, the authors use gCCA to separate different temporal sources in fMRI data. They assume that a set of common temporal responses to external stimulation is present in the subjects being studied, and show that they may be extracted using gCCA. In contrast, the underlying assumption in \cite{afshin2012enhancing} is that there are multiple subjects that share an unknown spatial response (or spatial map) to the common experimental excitation, but may show different temporal responses to external stimulation. Under the same assumption, the authors of \cite{varoquaux2010group} propose the application of gCCA for the estimation of a common spatial subspace that is spanned from the common, across subjects, spatial components, and use this estimate in order to form a preprocessing step before applying the ICA method. Finally, the estimation of ``common'' subspaces from multiple datasets, via CCA and gCCA based methods, has been considered in \cite{zhou2015group,ibrahim2019cell}.

A different line of research has been spanned by approaches based on the Deep Neural Network (DNN). For example, the authors of \cite{huang2017modeling} present a Deep Convolutional Autoencoder based approach, while in \cite{hu2018latent} the authors propose a Restricted Boltzmann Machine based approach. Moreover, Deep Recurrent Neural Network and Generative Adversarial Network based approaches have been proposed in \cite{li2021simultaneous} and \cite{dong2020novel}, respectively. The latter three works are able to decompose the fMRI data into different spatio-temporal components. However, none of these approaches can (1) identify which of the components are common or (2) determine how many common components exist in the dataset. To resolve these issues, these methods exploit prior information.

\subsection{Our Contribution}

We propose a new data generating model which takes into consideration both the common task-related spatial component and the common resting-state spatial components. We adopt the assumptions of \cite{afshin2012enhancing}, with respect to the common spatial maps, and assume the existence of one common temporal component, which is related to the common experimental excitation. We use gCCA and estimate the subspace that is spanned by the common spatial components, both task-related and resting-state. Based on this estimate, we perform a second gCCA and compute the common task-related temporal component. Finally, we use the estimated common task-related temporal component to compute an estimate of the associated common task-related spatial component and construct the respective activation map. The experimental tests of our method with synthetic data reveal that we are able to obtain very accurate temporal and spatial parameter estimates even at very low Signal to Noise Ratio (SNR). The tests with real-world fMRI data validate our data  model assumptions and show significant advantage of our approach over standard procedures based on General Linear Models (GLMs).

We note that an early version of this work has appeared in \cite{karakasis2020multi}. In this manuscript, we extend the work of \cite{karakasis2020multi} by proposing a data-driven method which effectively estimates the dimension of the common spatial subspace. Moreover, forming an estimate of the common subspace (as presented in \cite{karakasis2020multi}) and its dimension can be computationally demanding. In Appendix \ref{appendix:Compression}, we show that, under mild conditions, it is possible to overcome the computational bottleneck without affecting the quality of the resulting estimates. 

In order to demonstrate the efficiency of our method, we extend the experimental section of \cite{karakasis2020multi} by including experiments with synthetic data. The existence of ground truth in these cases enables us to test the effectiveness of our approach at each step, as well as in total. Furthermore, in this work, we provide a detailed comparison between the proposed approach and the conventional GLM, which enables us to highlight the differences between the two methods. At last, in this extended version, we apply our method to three additional real-world datasets; the results can be found in the Supplementary Material of this manuscript and they are in agreement with the results of the main part of the paper, demonstrating the effectiveness of
our approach.

\subsection{Notation}

Scalars, vectors, and matrices are denoted by small, small bold, and capital bold letters, for example, $x$, $\mathbf{x}$, $\mathbf{X}$. Sets are denoted by blackboard bold capital letters, for example, $\mathbb{U}$. $\mathbb{R}$ denotes the set of real numbers. $\mathbb{R}^{I\times J}$ denotes the set of ($I \times J$) real matrices. Inequality $\mathbf{X}\geq \mathbf{0}$ means that matrix $\mathbf{X}$ has nonnegative elements and $\mathbb{R}_+^{I\times J}$ denotes the set of ($I \times J$) real matrices with nonnegative elements. $\left\|{\bf x}\right\|_2$ 
denotes the Euclidean norm of vector ${\bf x}$, while
$\left\|{\bf X}\right\|_2$ and $\left\|{\bf X}\right\|_F$ denote, respectively, the spectral and the Frobenius norm of matrix ${\bf X}$. The transpose and the pseudoinverse of matrix $\mathbf{X}$ are denoted by $\mathbf{X}^T$ and $\mathbf{X}^{\dagger}$. The linear space spanned by the columns of matrix ${\bf X}$ is denoted by $\spn({\bf X})$. The orthogonal projector onto a linear subspace ${\cal S}$ is denoted by ${\bf P}_{\cal S}$. 
Finally, we use the Matlab-like expressions $\mathbf{X}(:,l)$ and $\mathbf{X}(k,:)$, which denote, respectively, the $l$-th column and the $k$-th row of matrix $\mathbf{X}$.

\section{fMRI data generating model}

We assume that the fMRI data have been arranged in matrices, where each matrix row
contains the time-series associated with a particular voxel. We denote these matrices as
$\left\{\mathbf{X}_k\right\}_{k=1}^K$, where $\mathbf{X}_k\in\mathbb{R}^{N\times M}$ denotes the data of the $k$-th subject, $N$ denotes the number of voxels and $M$ denotes the number of time points (note that, in general, $N\gg M$). Let $R$ be a positive integer smaller than $M$.

For each matrix $\mathbf{X}_k$, for $k=1,\ldots,K$, we adopt the model
\begin{equation}
\label{data_model}
\mathbf{X}_k =  \lambda_k \mathbf{a} \mathbf{s}^T+\mathbf{A}\mathbf{S}_k^{T} +\mathbf{E}_k,
\end{equation}
where
\begin{enumerate}
\item 
$\mathbf{a}\in\mathbb{R}_+^{N}$ and $\mathbf{s}\in\mathbb{R}^{M}$ denote, respectively, the common, to all subjects, task-related spatial and temporal component, and $\lambda_k\in \mathbb{R}_+$ denotes the intensity of the common rank-one term for the $k$-th subject;
\item 
$\mathbf{A}\in \mathbb{R}_+^{N\times (R-1)}$, whose columns are the common, to all subjects, spatial components related with the spontaneous fMRI activity;
\item 
$\mathbf{S}_k\in\mathbb{R}^{M\times (R-1)}$, whose columns are the temporal components associated with the spontaneous fMRI activity and, in general, vary across subjects. We assume that 
\begin{equation}
\bigcap_{k=1}^K \spn\left(\mathbf{S}_k\right) = \emptyset,
\label{assumption_empty_intersection}
\end{equation}
that is, there is no subspace that is common to {\em all} $\spn\left(\mathbf{S}_k\right)$, for $k\in\{1,\ldots, K\}$ (we shall investigate the validity of this important assumption later); 

\item 
$\mathbf{E}_k \in \mathbb{R}^{N\times M}$ denotes the ``unmodelled fMRI signal'' of the $k$-th subject and is considered as (strong) additive noise. We assume that terms ${\bf E}_k$, for $k=1,\ldots,K$, are statistically independent from each other. 
\end{enumerate}
We propose model (\ref{data_model}) based on both the existing literature \cite{fox2007spontaneous,fair2007method,fox2006coherent,greicius2004default,arfanakis2000combining,fransson2006default,erhardt2011comparison
,afshin2012enhancing,calhoun2001spatial} and the detailed examination of our real-world data. 

We note that, in \cite{karakasis2020multirest}, where we consider the resting-state scenario, we have assumed that  matrix ${\bf A}$ is nonnegative and orthogonal, implying a brain parcellation structure. 
However, in this work, our focus is on the task-related scenario and this assumption is not necessary.

Our aim is to obtain an accurate estimate of the common spatial term ${\bf a}$, which will lead to a precise activation brain map and, thus, to the accurate localization of the stimulated brain areas.

In order to use simpler notation, we define the matrix of the common spatial components 
\begin{equation}
\mathbf{W}:=\left[\mathbf{a}~\mathbf{A}\right]\in\mathbb{R}_+^{N\times R},
\label{W_def}
\end{equation}
and the matrices of the temporal components 
\begin{equation}
\mathbf{Z}_k := \left[\lambda_k\mathbf{s}~\mathbf{S}_k\right]\in\mathbb{R}^{M\times R}, \quad
\mbox{for}~k=1,\ldots,K.
\label{Zk_def}
\end{equation}
We further assume that matrices ${\bf W}$ and 
${\bf Z}_k$, for $k=1,\ldots,K$, are full-column rank. Using this notation, matrix ${\bf X}_k$, defined in 
(\ref{data_model}), can be expressed as
\begin{equation}
{\bf X}_k = {\bf W} {\bf Z}_k^T + {\bf E}_k.
\label{data_model_1}
\end{equation}

\section{Estimation of the common spatial and temporal components}

Our approach towards the construction of the task-related brain excitation map consists of three stages:
\begin{enumerate}
\item 
we use ${\bf X}_k$, for $k=1,\ldots,K$, and obtain an orthonormal basis for an estimate of the common spatial subspace, $\spn({\bf W})$,
by solving a gCCA problem;

\item 
we use the solution of the first stage and estimate the unique common time component, ${\bf s}$,
by solving a second gCCA problem;

\item 
we use the estimate of ${\bf s}$, and obtain an estimate of ${\bf a}$, which is our final target.
\end{enumerate}

\subsection{Estimation of the common spatial subspace}
\label{subsection_common_spatial_CCA}

We assume that the dimension, $R$, of the common spatial subspace, $\spn({\bf W})$, is known (later,
we shall present a criterion for the estimation of $R$ from the data).

In order to estimate an orthonormal basis for the common
spatial subspace, $\spn({\bf W})$, we adopt the MAX-VAR formulation of the gCCA \cite{horst1961generalized}
and solve the optimization problem
\begin{equation}
\begin{split}
& \min_{\{\mathbf{Q}_k\}_{k=1}^K,\mathbf{G}}
\sum_{k=1}^K\left\|\mathbf{X}_k\mathbf{Q}_k-\mathbf{G}\right\|_F^2\\
&\quad \text{subject to}~\mathbf{G}^T\mathbf{G}=\mathbf{I}_R,
\label{common_spatial_gCCA}
\end{split}
\end{equation}
where ${\bf Q}_k\in\mathbb{R}^{M\times R}$, for $k=1,\ldots,K$, and
${\bf G}\in\mathbb{R}^{N\times R}$. 

The solution ${\bf Q}_k^o$, for $k=1,\ldots,K$, and ${\bf G}^o$ 
of problem (\ref{common_spatial_gCCA}) can be computed as follows. 
For a fixed ${\bf G}$, the optimal $\mathbf{Q}_k$ can be expressed as 
\begin{equation}
\mathbf{Q}_k({\bf G})=\mathbf{X}_k^{\dagger}\mathbf{G}, ~\mbox{for}~k=1,\dots,K.
\label{Q_k_def}
\end{equation} 
If we substitute this value into (\ref{common_spatial_gCCA}), then the problem becomes
\begin{equation}
\underset{\mathbf{G}^T\mathbf{G}=\mathbf{I}_R} \max \, \text{Tr}\left(\mathbf{G}^T\left(\sum_{k=1}^K\mathbf{X}_k\mathbf{X}_k^{\dagger}\right)\mathbf{G}\right).
\end{equation}
We define
\begin{equation}
\mathbf{M} := \sum_{k=1}^K\mathbf{X}_k\mathbf{X}_k^{\dagger},
\end{equation}
with eigenvalue decomposition given by
$\mathbf{M} = {\bf U}_{M} \boldsymbol{\Lambda}_{M} {\bf U}_{M}^T$.
An optimal solution $\mathbf{G}^o$ is given by \cite{fu2017scalable}
\begin{equation} 
\mathbf{G}^o=\mathbf{U}_M\left(:,1:R\right).
\end{equation}
The fact that MAX-VAR gCCA indeed identifies the common subspace of the views in the noiseless case has been proved in \cite{ibrahim2019cell} and \cite{sorensen2021generalized} for the two view and multiview cases, respectively.
Notice that $\mathbf{M}$ is a $N\times N$ matrix. Hence, for the case of whole brain data analysis (where
$N$ is very large), 
the formation of matrix $\mathbf{M}$ and the computation of its eigenvalue decomposition may be 
prohibitive. In Appendix \ref{appendix:Compression}, we show that, if $KM\ll N$, then we can compress matrices $\left\{\mathbf{X}_k\right\}_{k=1}^K$ and efficiently
compute matrix $\mathbf{G}^o$, bypassing the computation of $\mathbf{M}$.

If the fMRI data matrices ${\bf X}_k$ were noiseless, that is, if ${\bf E}_k={\bf 0}$, for $k=1,\ldots,K$, then the solution of problem (\ref{common_spatial_gCCA}) would result to ${\bf G}^o$ such that (see (\ref{data_model_1})) 
\begin{equation}
\spn({\bf G}^o) = \spn({\bf W}).
\label{equality_span_Go}
\end{equation}
This implies that 
\begin{equation}
\mathbf{W}=\mathbf{G}^o\mathbf{P},
\end{equation}
for some $(R\times R)$ invertible matrix ${\bf P}$. Furthermore, in this case and for all $k\in\{1,\ldots,K\}$, matrices ${\bf Q}_k^o$ and ${\bf Z}_k$ would span the same subspace, namely, 
\begin{equation}
\spn({\bf Q}_k^o) = \spn\left({\bf Z}_k\right).
\label{equality_span_Qko} 
\end{equation}
This holds because
\begin{equation}
\begin{split}
{\bf Q}_k^o & =\mathbf{X}_k^{\dagger}\mathbf{G}^o=\left( \mathbf{Z}_k^{T}\right)^{\dagger} \mathbf{W}^{\dagger}\mathbf{G}^o=\mathbf{Z}_k\mathbf{F}, 
\end{split}
\end{equation}
where 
\begin{equation}
\mathbf{F}:=\left(\mathbf{Z}_k^T\mathbf{Z}_k\right)^{-1}\mathbf{P}^{-1}.
\label{F_def}
\end{equation}
The fact that ${\bf E}_k$, for $k=1,\ldots,K$, are nonzero makes (\ref{equality_span_Go}) and 
(\ref{equality_span_Qko}) {\em approximate}\/ equalities.

\subsection{Estimation of the common temporal component}
\label{subsection_common_time_component_extraction}

Having computed ${\bf G}^o$, we proceed to the estimation of ${\bf s}$ by assuming that 
(\ref{equality_span_Qko}) is exact. We shall test the accuracy of our assumption and 
the effectiveness of our approach in the section with the experimental results. 

Based on assumption (\ref{assumption_empty_intersection}) and definition 
(\ref{Zk_def}), we have that, in the noiseless case,  
\begin{equation}
\bigcap_{k=1}^K \spn\left(\mathbf{Z}_k\right) = \spn\left(\mathbf{s}\right),
\label{common_time_component}
\end{equation}
which, using (\ref{equality_span_Qko}), leads to 
\begin{equation}
\bigcap_{k=1}^K \spn\left(\mathbf{Q}_k^o\right) = \spn\left(\mathbf{s}\right).
\label{equality_intersection_span_Qko}
\end{equation}
We obtain an estimate of ${\bf s}$ by solving the MAX-VAR problem 
\begin{equation}
\begin{split}
& \min_{\{\mathbf{d}_k\}_{k=1}^K,\mathbf{g}} 
\sum_{k=1}^K\left\|\mathbf{Q}_k^o\mathbf{d}_k-\mathbf{g}\right\|_2^2\\
& \qquad \text{subject to}~\left\|{\bf g}\right\|_2=1.
\end{split}
\label{MAXVAR}
\end{equation}
If we denote the optimal ${\bf g}$ in (\ref{MAXVAR}) by ${\bf g}^o$, we have that 
\begin{equation}
\mathbf{g}^o = \pm \, \frac{\mathbf{s}}{\|\mathbf{s}\|_2}.
\label{equality_go_s}
\end{equation}
Since (\ref{equality_span_Qko}) defines a family of approximate equalities, equalities (\ref{equality_intersection_span_Qko}) and  (\ref{equality_go_s}) are approximate.

\subsection{Estimation of the common spatial component}
\label{subsection_estimation_of_a}

Until now, we have obtained the estimate of an orthonormal basis of the common spatial subspace, ${\bf G}^o$, and 
the estimate of the common temporal component, ${\bf g}^o$. We can proceed to the estimation of the common spatial component, ${\bf a}$, and the intensity vector, $\boldsymbol\lambda$, by following various paths.
A simple approach is to consider the problem
\begin{equation}
\underset{{\bf a}\ge {\bf 0}, {\boldsymbol \lambda}\ge {\bf 0}} \min \sum_{k=1}^K \|{\bf X}_k - \lambda_k {\bf a} {\bf g}^{o^T}\|_F^2.
\label{Prob_1}
\end{equation}
However, in this case, we do not exploit the fact that ${\bf a}\in\spn(\mathbf{G}^o)$. We can enforce this constraint
if we compute the projected data 
\begin{equation}
\mathbf{X}^o_k :=\mathbf{P}_{\spn\left(\mathbf{G}^o\right)}\mathbf{X}_k, \quad k=1,\dots,K,
\label{X_ko_def}
\end{equation}
and solve the problem
\begin{equation}
\underset{{\bf a}\geq{\bf 0},{\boldsymbol \lambda}\geq {\bf 0}} \min \, \sum_{k=1}^K \|{\bf X}^o_k - \lambda_k {\bf a} {\bf g}^{o^T}\|_F^2,
\label{Prob_2}
\end{equation}
whose solution, $({\bf a}^o, {\bf \boldsymbol\lambda}^o)$, is our final estimate of  $({\bf a},\boldsymbol\lambda)$.

\subsection{On the Dimension of Common Subspaces}
\label{Subsection_On_Dimension_of_Common_Subspaces}

In Subsection \ref{subsection_common_spatial_CCA}, we assumed that we know the true dimension, $R$, of the common spatial subspace, $\spn({\bf W})$, and derived the estimate ${\bf G}^o$ of an  orthonormal basis of $\spn({\bf W})$. Of course, in general, the value of $R$ is unknown, thus, we must estimate it from the data. 
In the sequel, we provide a procedure which gives us very useful information about the value of $R$ \cite{karakasis2020multirest}. We note that the same procedure can be used for the verification of assumption (\ref{assumption_empty_intersection}).

Let the hypothesized dimension of the common spatial subspace be denoted by $\hat{R}$. At first, we assume that 
$\hat{R}=R$.
Let ${\cal K}_1$ and ${\cal K}_2$ be a random partition of the set of the subjects $\mathcal{K}=\{1,\ldots,K\}$. In the noiseless case, if we solve problem 
(\ref{common_spatial_gCCA}) twice, for $k\in{\cal K}_1$ and $k\in{\cal K}_2$, and call the resulting orthonormal bases ${\bf G}_1^o$ and ${\bf G}_2^o$, respectively, then it holds true that
\begin{equation}
\spn({\bf G}_1^o) = \spn({\bf G}_2^o).
\end{equation}
That is, the common spatial subspaces coincide. 
If we start adding noise and repeat the process, then 
the estimated subspaces, $\spn({\bf G}_1^o)$ and $\spn({\bf G}_2^o)$, will start to move away from each other.
One way to measure the distance between a pair of linear subspaces ${\cal S}_1$ and 
${\cal S}_2$ is to compute their gap, defined as 
\cite[p. 93]{stewart1990computer} 
\begin{equation}
\begin{split}
\rho_{g,2}\left({\cal S}_1,{\cal S}_2\right) & := \| {\bf P}_{{\cal S}_1} - {\bf P}_{{\cal S}_2}\|_2.
\end{split}
\end{equation}
If $\hat{R}=R$ and $\|{\bf E}_k\|_2=O(\epsilon)$, for $k=1,\ldots,K$, where $\epsilon$ is a small positive number, then we expect that 
\begin{equation}
\| {\bf P}_{\spn({\bf G}_1^o)} - {\bf P}_{\spn({\bf G}_2^o)} \|_2 =
O(\epsilon).
\end{equation}
If $\hat{R}>R$, then, if we solve (\ref{common_spatial_gCCA}) for
${\cal K}_1$ and ${\cal K}_2$, then, besides the $R$-dimensional common subspace, $\spn({\bf W})$, we try to model ``common'' noise subspace. Since the noise terms ${\bf E}_k$ are statistically independent across subjects and $N\gg M$, we do not expect to find any common noise subspace in the datasets associated with ${\cal K}_1$ and ${\cal K}_2$. Thus, if $\hat{R}>R$, we expect that 
\begin{equation}
\| {\bf P}_{\spn({\bf G}_1^o)} - {\bf P}_{\spn({\bf G}_2^o)} \|_2 \approx 1.
\end{equation}
Finally, if $\hat{R}<R$ and the rank-one terms which constitute the products ${\bf W}{\bf Z}_k^T$ are of almost ``equal strength,'' then we expect that 
\begin{equation}
O(\epsilon) \lesssim \| {\bf P}_{\spn({\bf G}_1^o)} - {\bf P}_{\spn({\bf G}_2^o)} \|_2 \lessapprox 1,
\end{equation}
because $\spn({\bf G}_1^o)$ and $\spn({\bf G}_2^o)$  
will ``randomly'' capture $\hat{R}$ out of $R$ dimensions of the common spatial subspace.

Thus, the gap between ${\rm col}({\bf G}_1^o)$ and 
${\rm col}({\bf G}_2^o)$ provides valuable information about the true dimension of the common subspace, ${\rm col}({\bf W})$.
Accurate expressions for the gap lie beyond the scope of this manuscript, require tools from matrix perturbation theory, and pose stringent assumptions on the size of the noise, which may not be fulfilled in our case. We shall test the usefulness of our claims in Subsection \ref{Subsection_Real_World_Data}.

\section{Experiments}
\label{Section_Experiments}

\subsection{Synthetic Data}
\label{Subsection_Synthetic_Data}

First, we test the effectiveness of our approach using synthetic data. 
We remind that our main goal is the estimation of the common task-related spatial component, 
${\bf a}$. We generate random data 
$\{{\bf X}_k\}_{k=1}^K$
according to the model
\begin{equation}
\mathbf{X}_k = \lambda_k\mathbf{a}\mathbf{s}^T + \beta\left(\mathbf{A}\mathbf{S}^T_k+\mathbf{E}_k\right),
\end{equation}
where matrices $\left\{\beta\mathbf{A}\mathbf{S}^T_k\right\}_{k=1}^K$ and $\left\{\beta\mathbf{E}_k\right\}_{k=1}^K$ act as noise terms. We define the SNR as
\begin{equation}
\text{SNR} := \frac{\sum_{k=1}^K\left\|\lambda_k\mathbf{a}\mathbf{s}^T\right\|_F^2}{\sum_{k=1}^K \beta^2 \left\|\mathbf{A}\mathbf{S}^T_k+\mathbf{E}_k\right\|_F^2}.
\end{equation}
Of course, the relative power of terms ${\bf A}{\bf S}_k$ and ${\bf E}_k$, for $k=1,\ldots,K$,
is very important. 
The values of the parameters used in our experiments are $N=10^5$, $M=100$, $K=25$, and $R=30$. 
Vectors $\mathbf{a}$ and $\boldsymbol{\lambda}$, as well as matrix $\mathbf{A}$, have 
independent and identically distributed (i.i.d.) elements, taking values uniformly at random in the interval [0,1]. Vector $\mathbf{s}$ and matrices $\mathbf{S}_k$, for $k=1,\ldots,K$, have i.i.d. elements following the normalized Gaussian distribution, $\mathcal{N}\left(0,1\right)$. On the other hand, matrices $\mathbf{E}_k$, for $k=1,\ldots,K$, have i.i.d. elements, ${\cal N}(0,\sigma_{\bf E}^2)$, with $\sigma^2_{\bf E}$ chosen such that
\begin{equation}
\frac{\sum_{k=1}^K\|{\bf A}{\bf S}_k^T\|_F^2}{\sum_{k=1}^K\|{\bf E}_k\|_F^2}=c,
\end{equation}
where $c$ is a given positive real number. After applying our method to the real-world fMRI datasets examined in subsection
\ref{Subsection_Real_World_Data}, we were able to estimate all the factors that appear in relation ($28$). Based on these estimates, we observed that the value of $c$ was approximately equal to $0.33$ for all the considered real-world datasets. This observation motivated us to use this value of $c$ in our experiments with synthetic datasets.

We fix $\boldsymbol{\lambda}$, $\mathbf{a}$, and $\mathbf{s}$, and compute the mean (over $100$ independent realizations of $\mathbf{A}$, $\mathbf{S}_k$, and $\mathbf{E}_k$,  for $k=1,\ldots,K$) correlation coefficients between the true and the estimated quantities. In the remaining of this section, the fixed components $\boldsymbol{\lambda}$, $\mathbf{a}$, and $\mathbf{s}$ are denoted as $\boldsymbol{\lambda}_{\rm true}$, $\mathbf{a}_{\rm true}$, and $\mathbf{s}_{\rm true}$, respectively. 

\begin{figure}
\center
\includegraphics[scale=0.5]{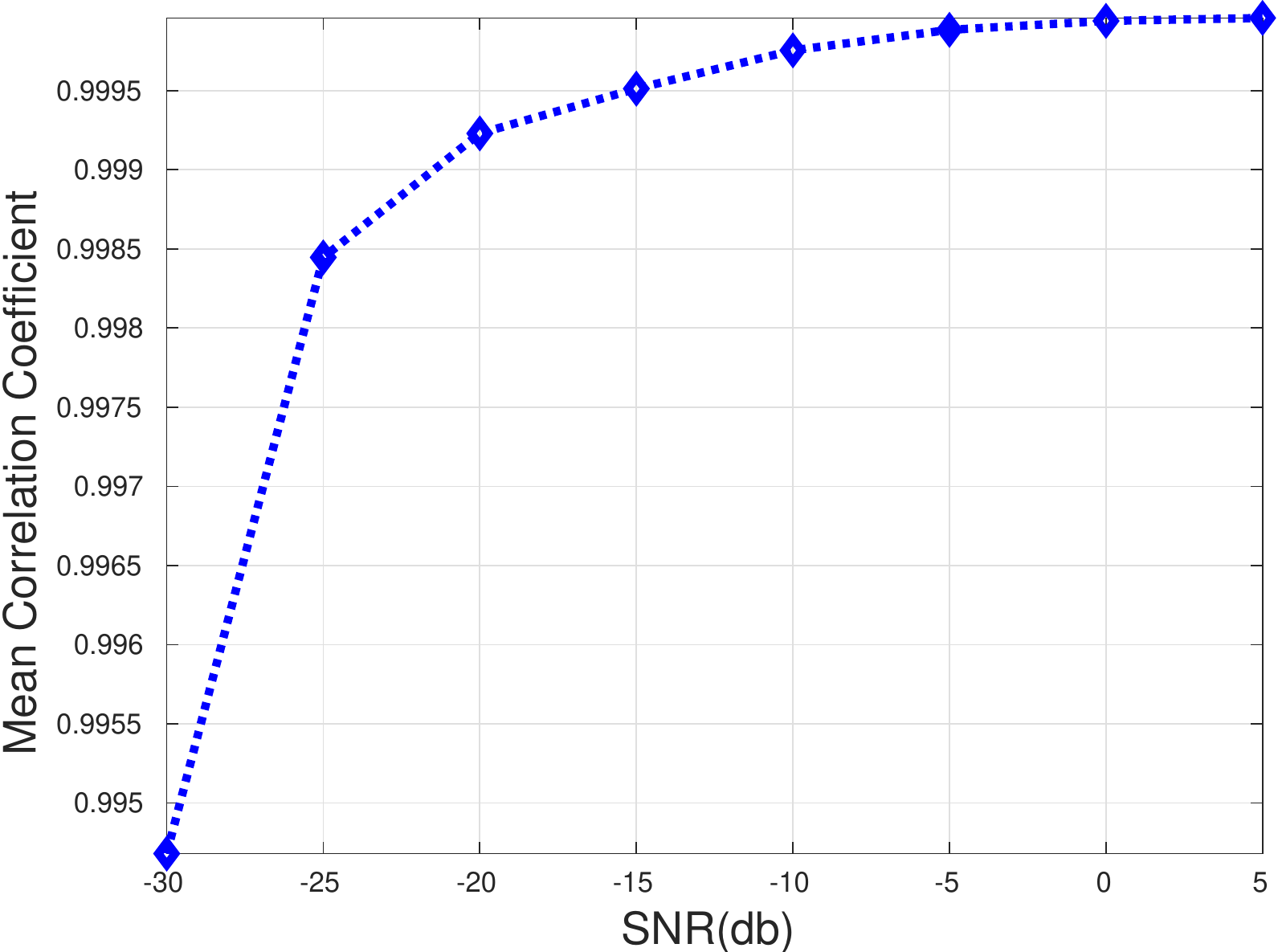}
\caption{Mean correlation coefficient between $\mathbf{s}_{\rm true}$ and $\mathbf{s}_{\rm est}$ versus SNR.} 
\label{fig:Exp_Res_s_est}
\includegraphics[scale=0.5]{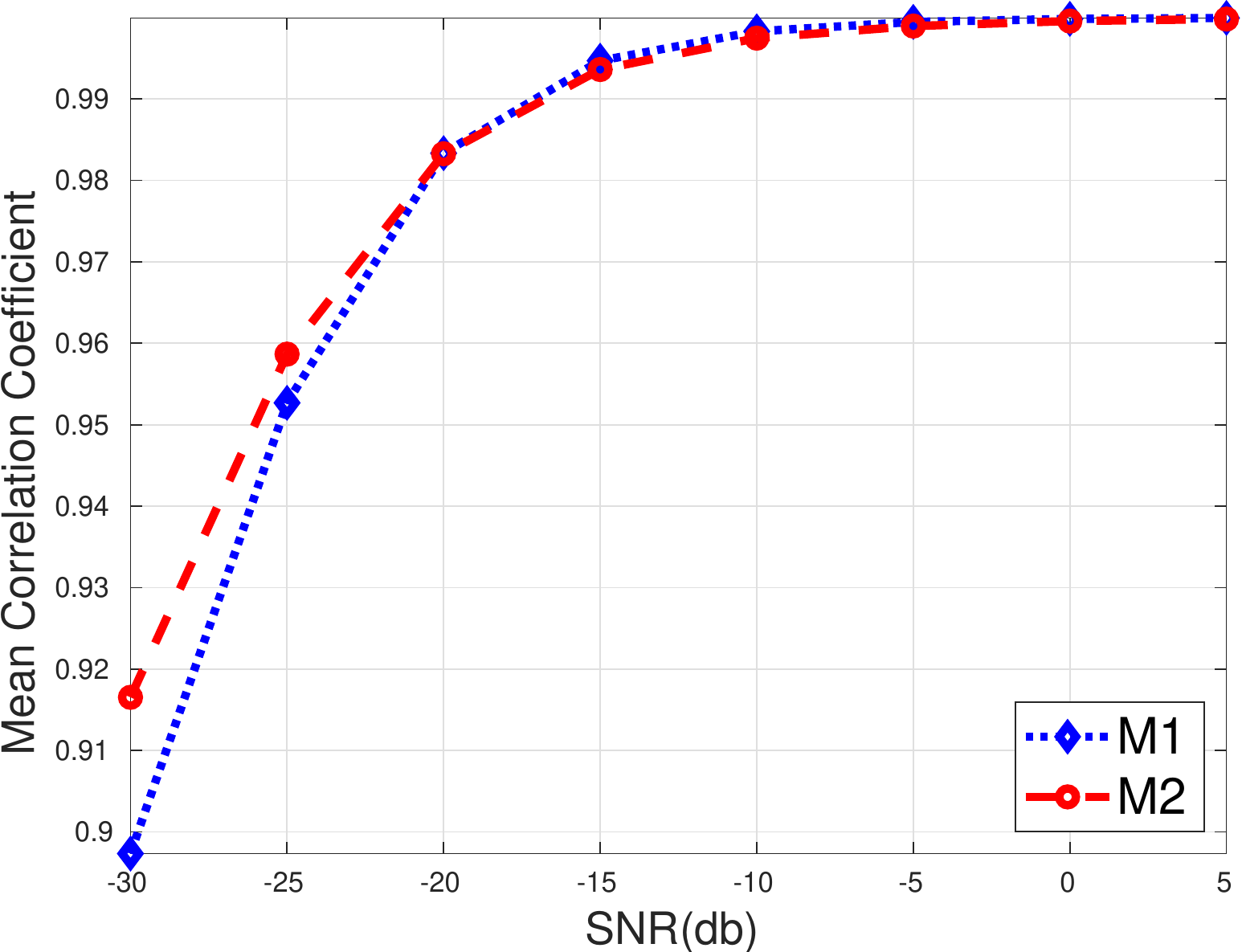}
\caption{Mean correlation coefficient between $\mathbf{a}_{\rm true}$ and $\mathbf{a}_{\rm est}$ versus SNR, for methods M$1$ and M$2$.}
\label{fig:Exp_Res_a_est}
\includegraphics[scale=0.5]{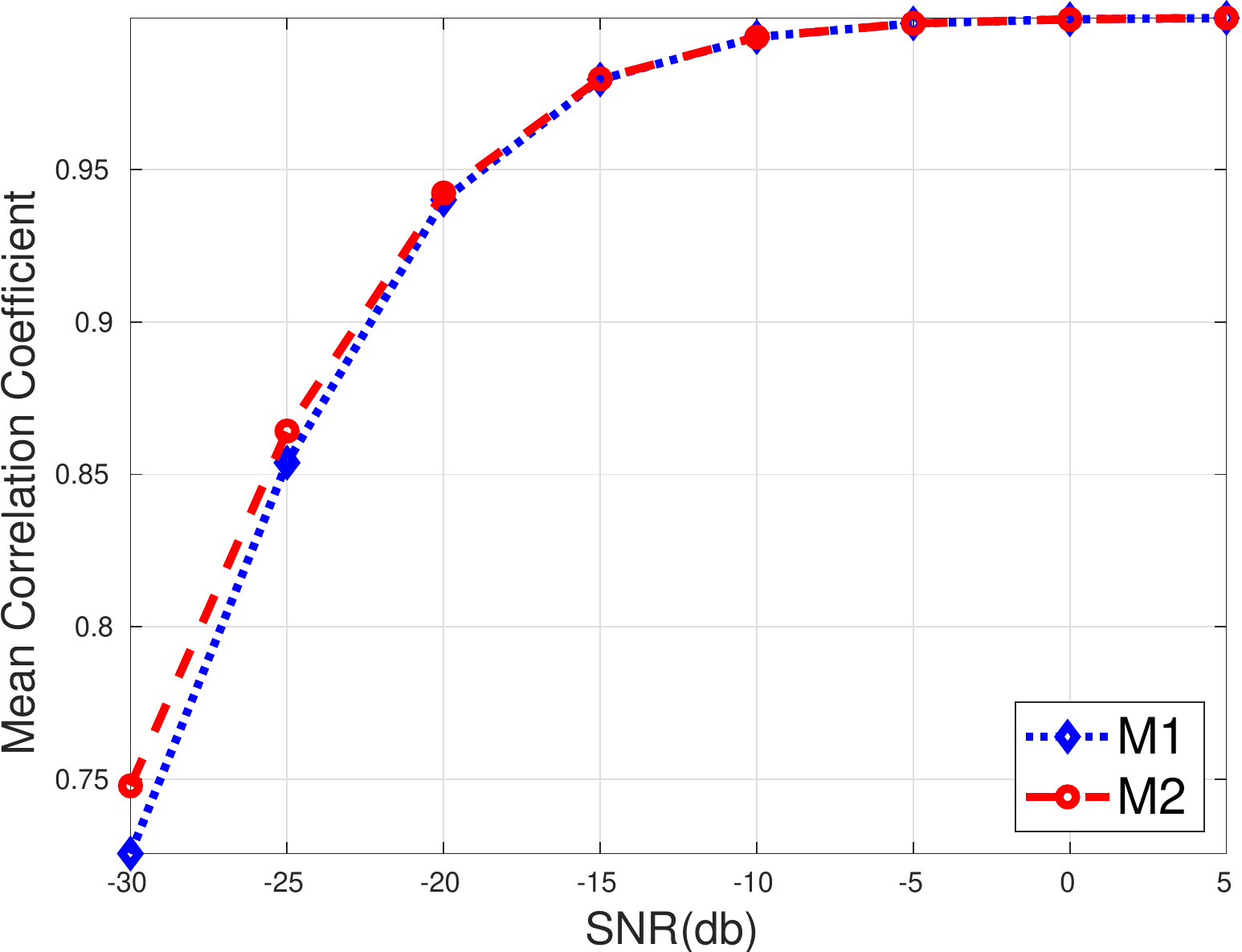}
\caption{Mean correlation coefficient between ${\boldsymbol\lambda}_{\rm true}$ and $\boldsymbol{\lambda}_{\rm est}$ versus SNR, for methods M$1$ and M$2$.}
\label{fig:Exp_Res_lambda_est}
\end{figure}

At first, we test the accuracy of our estimate of the common temporal component, ${\bf s}$. We perform the two-stage gCCA and compute our estimate, ${\bf s}_{\rm est}$, via (\ref{MAXVAR}). In Fig. \ref{fig:Exp_Res_s_est},
we depict the accuracy of our estimate versus SNR. We observe that we attain very high estimation
accuracy even at SNR as low as $-30$\,db.

Then, we use ${\bf s}_{\rm est}$ and test the accuracy of our estimates of ${\bf a}$ and $\boldsymbol{\lambda}$ computed by solving problems (\ref{Prob_1}) and (\ref{Prob_2}). 
We denote the methods that are based on solving problems (\ref{Prob_1}) and (\ref{Prob_2}) as method $1$ (M$1$) and method $2$ (M$2$), respectively. We solve these problems via an Alternating Optimization (AO) approach. Since AO is sensitive to local minima, we solve the problems for $5$ random initializations and take into account only the solution attaining the best fit. We depict the results in Figs. \ref{fig:Exp_Res_a_est} and \ref{fig:Exp_Res_lambda_est}, where we observe that M$2$ outperforms M$1$ for very low SNR ($-30$\,db to $-20$\,db). 

 
In summary, we observe that our approach attains very accurate estimates even at very low SNR, which is usually the case with fMRI data processing.

\subsection{Real-World Data}
\label{Subsection_Real_World_Data}

In this subsection, we test our approach using real-world task-related fMRI data. More specifically, we process four datasets, recorded at the University of Crete General Hospital, from a group of $25$ healthy adults performing four visual tasks, which were identical in all but one aspect (the precise kinematics of an observed person-directed action). Next, we quote a short description of the experiment design and the preprocessing pipeline that was applied to the data. A more extensive description of the experiments can be found in the Supplementary Material. Then, we present the results obtained by analyzing the data using our method.

\subsubsection{Experiment design}

The fMRI block design consists of four action observation conditions, each involving four ``active'' $35$\,sec blocks alternating with four $35$\,sec baseline blocks. Within each ``active'' block, a video clip illustrating a two-movement action sequence was presented $6$ times, while the stimulus set-up was identical across blocks and conditions.

The data employed in the main analysis reported here were derived from the first experimental condition (or, briefly, condition (i)), examining the effects of an action with the same goal but different kinematics. 
The results concerning the other three experimental conditions are presented in the Supplementary Material.

\subsubsection{Image acquisition and pre-processing}

Scanning was performed on an upgraded $1.5$\,T Siemens Vision/Sonata scanner (Erlangen, Germany) with powerful gradients (Gradient
strength: $45$\,mT/m, Gradient slew rate: $200$\,mT/m/ms) and a standard four channel head array coil. For the BOLD-fMRI, a $T2^*$-weighted, fat-saturated $2D$-FID-EPI sequence was used with the following parameters: repetition time (TR) $3500$\,ms, echo time (TE) $50$\,ms, field of view (FOV) $192 \times 192 \times 108~(x, y, z)$, acquisition voxel size $3 \times 3 \times 3$\,mm. Whole brain scans consisted of $36$ transverse slices with $3.0$-mm slice thickness and no interslice gap. The time-series recorded in each condition comprised $80$ volumes (time points). 
In our analysis, we ignore the first $5$ volumes of each time-series, as is customary in fMRI studies.

Image preprocessing was performed in SPM8.\footnote{Statistical Parametric Mapping software, SPM: Welcome Department
of Imaging Neuroscience, London, UK; available at: http://www.fil.ion.ucl.ac.uk/spm/.} Initially, EPI scans were spatially realigned to the first
image of the first time-series using second-degree B-spline interpolation algorithms and motion-corrected through rigid body transformations.
Next, images were spatially normalized to a common brain space (MNI template) and smoothed using an isotropic Gaussian filter (FWHM=$8$\,mm). At last, all voxel time series, from all subjects, were centered and de-drifted (subtraction of the mean value and linear terms).

We note that the SPM platform is able to provide a time response component, based on the activation onsets and offsets, which is expected to appear in the activated brain voxels. This response is the same for all four experimental conditions since, as we mentioned, the stimulus layout, in all conditions, is the same. From now on, we denote this response as $\mathbf{s}_{\rm exp}$.

\begin{figure}
\center
\includegraphics[width=0.9\linewidth]{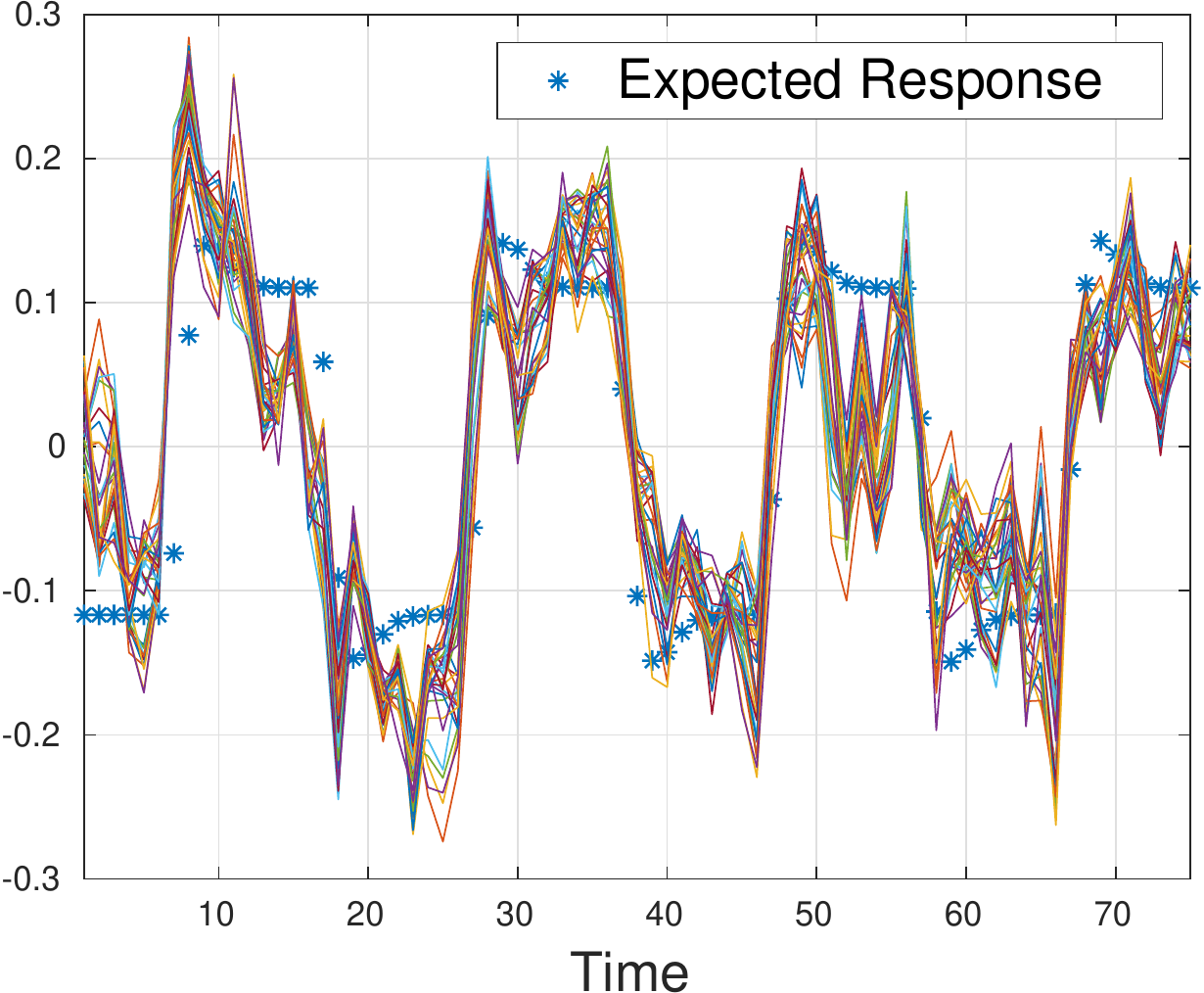}
\caption{Estimate $\mathbf{s}_{\rm est}$ for condition (i) and varying common subspace dimensions, from $10$ to $40$. The signal depicted with blue stars is the $\mathbf{s}_{\rm exp}$.\vspace{2mm}}
\label{fig:1a}
\includegraphics[width=0.9\linewidth]{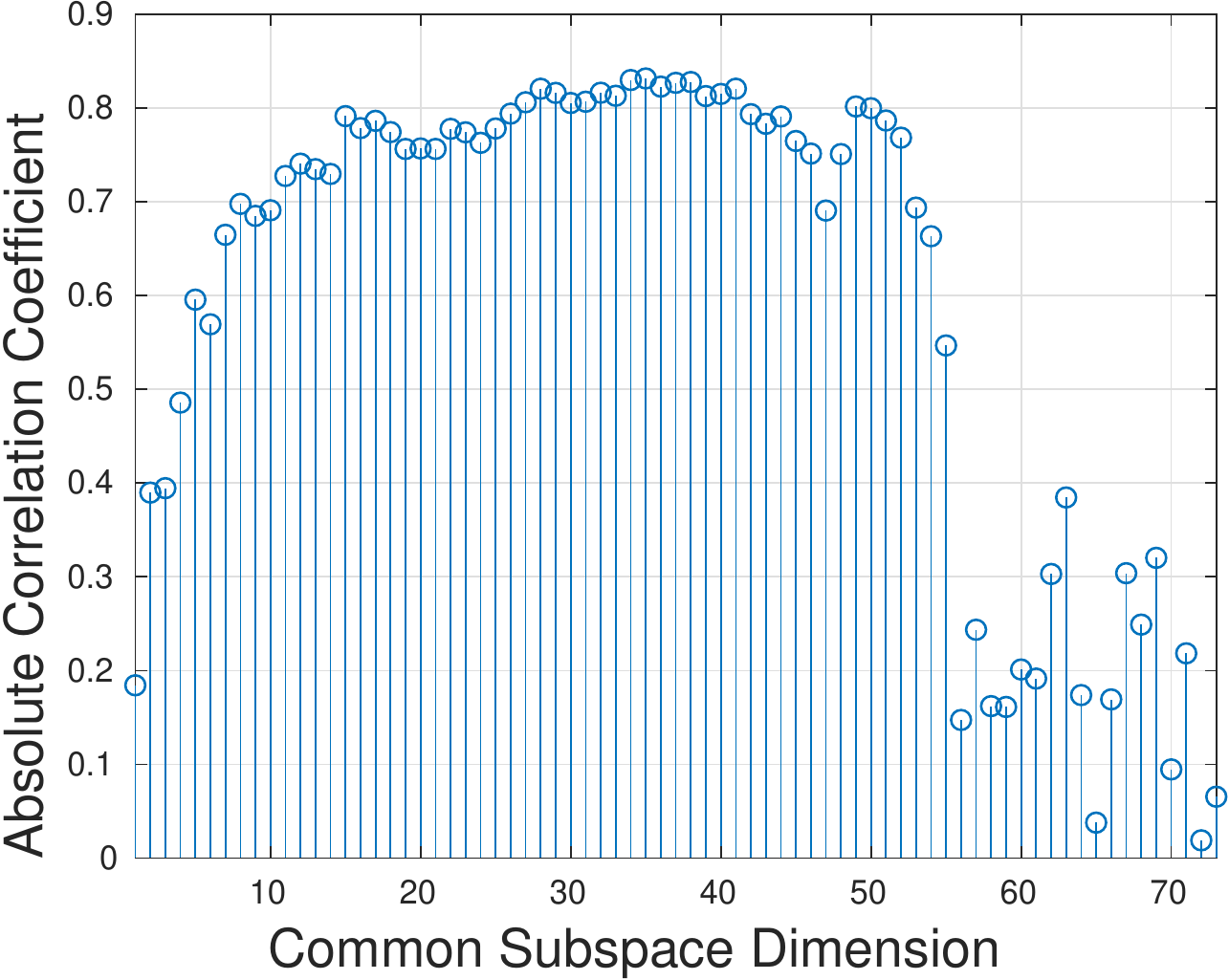}
\caption{ Absolute correlation coefficient between $\mathbf{s}_{\rm exp}$ and $\mathbf{s}_{\rm est}$ across different common subspace dimensions, for condition (i).\vspace{2mm}}
\label{fig:2a}
\includegraphics[width=0.9\linewidth]{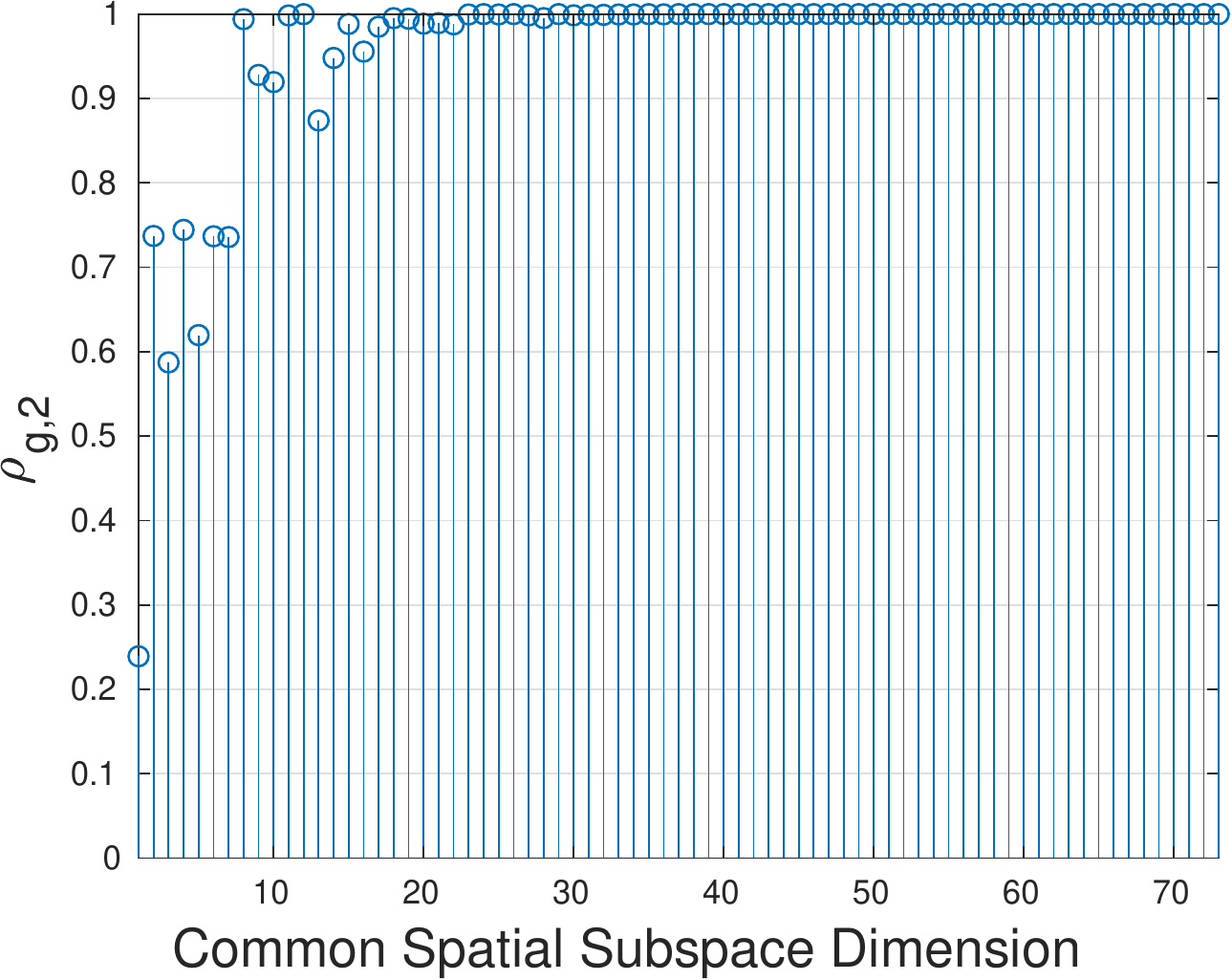}
\caption{Gap function $\rho_{g,2}\left({\cal S}_1,{\cal S}_2\right)$ evaluated for varying dimension of spatial subspaces ${\cal S}_1$ and ${\cal S}_2$, for condition (i).\vspace{6mm} } 
\label{fig:2lpa}
\end{figure}

\begin{figure}
\center
\includegraphics[width=0.9\linewidth]{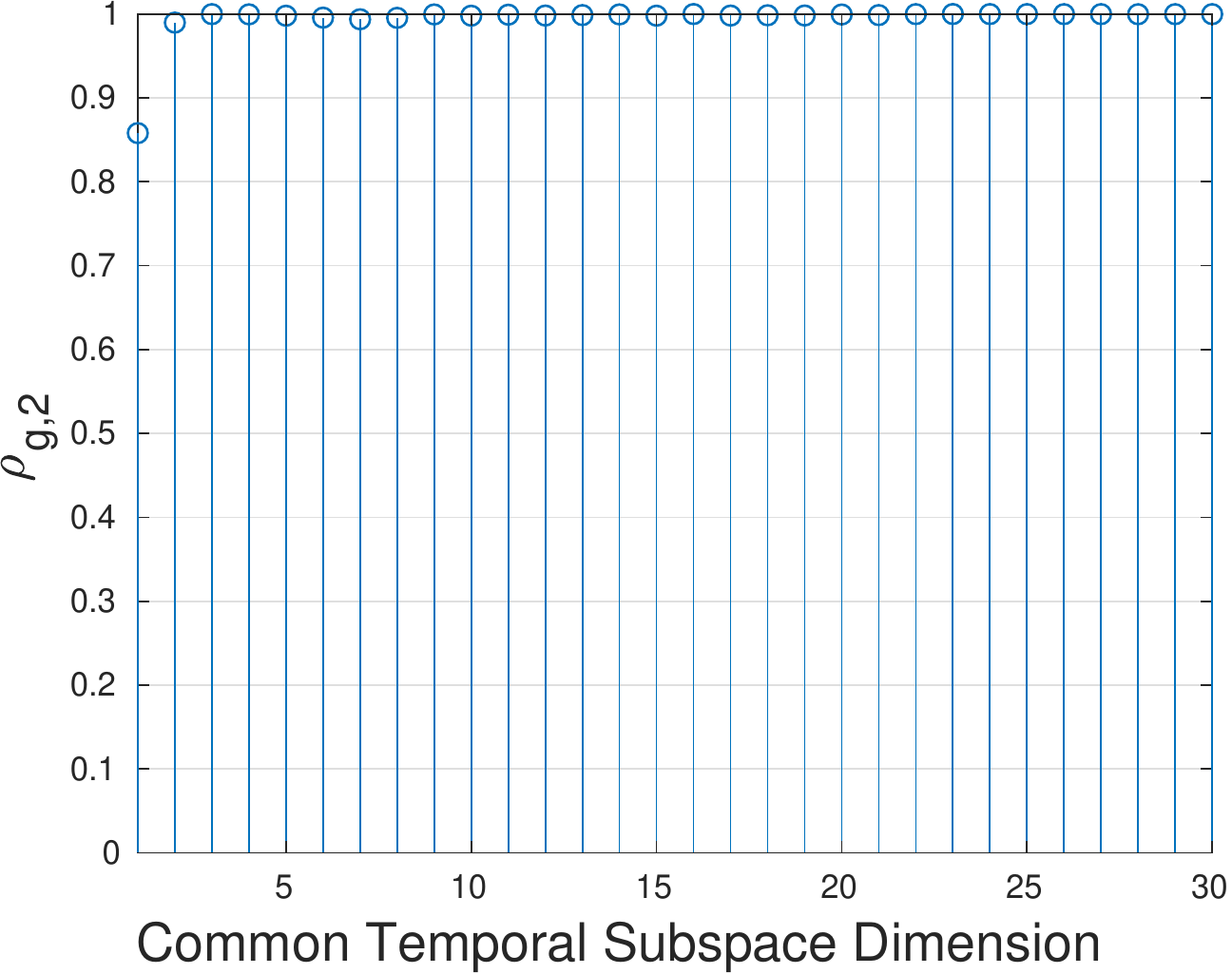}
\caption{Gap function $\rho_{g,2}\left({\cal T}_1,{\cal T}_2\right)$ evaluated for spatial subspace of dimension equal to $30$ and varying dimension of temporal subspaces ${\cal T}_1$ and ${\cal T}_2$, for condition (i). \vspace{2mm} } 
\label{fig:2ka}
\centering
\includegraphics[width=0.9\linewidth]{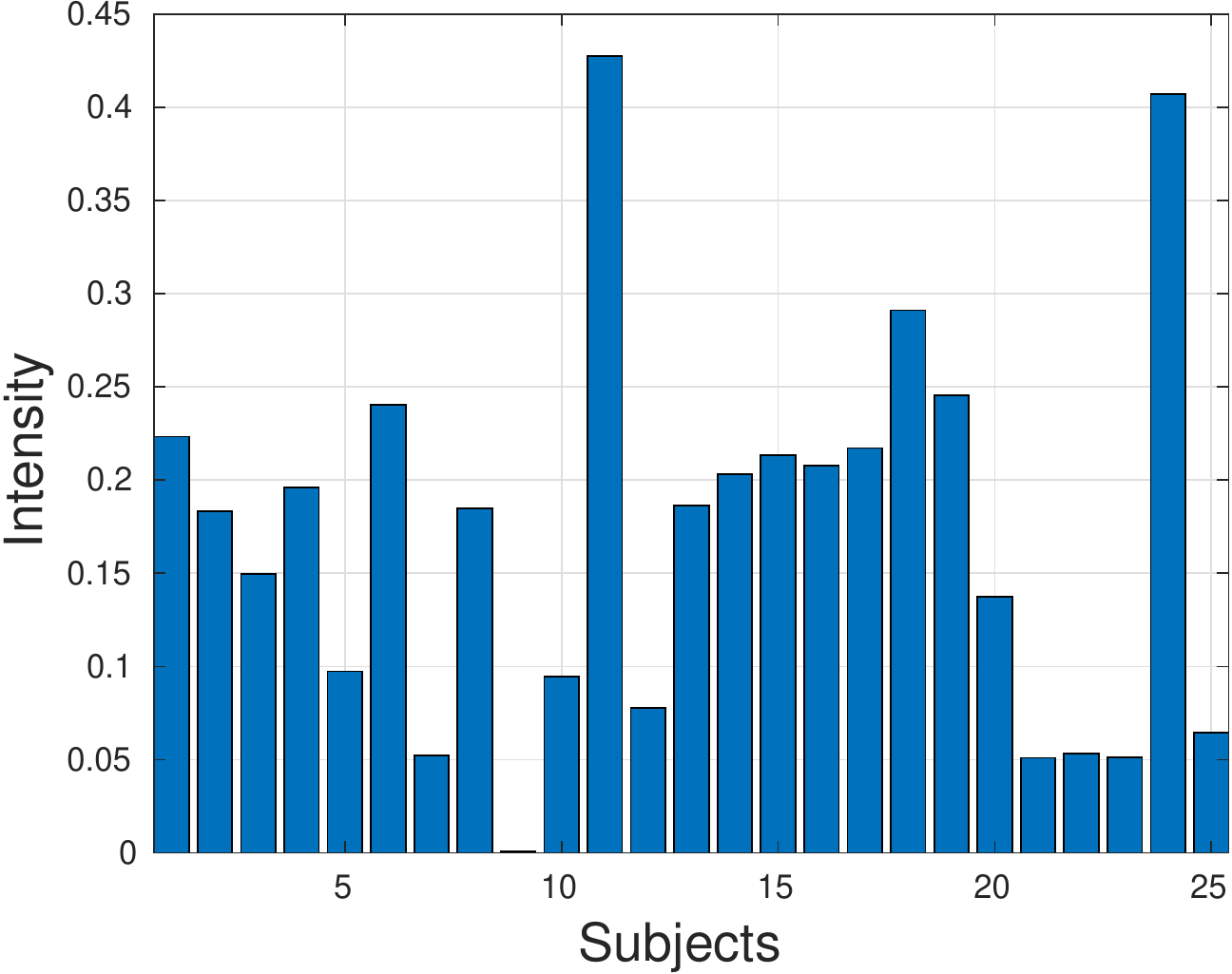}
\caption{Estimate $\boldsymbol{\lambda}_{\rm est}$ of vector of intensities for condition (i) and ``common'' subspace dimension equal to $30$.\vspace{2mm} } 
\label{fig:2lqa}
\end{figure}

\begin{figure}
\centering
\begingroup
\sbox0{\includegraphics{example-image}}
\makeatletter
\Gscale@div\myscale{9.25cm}{\wd0}
\includegraphics[clip,trim={1.56\wd0 0.35\wd0 1.30\wd0 0.335\wd0},scale=\myscale]{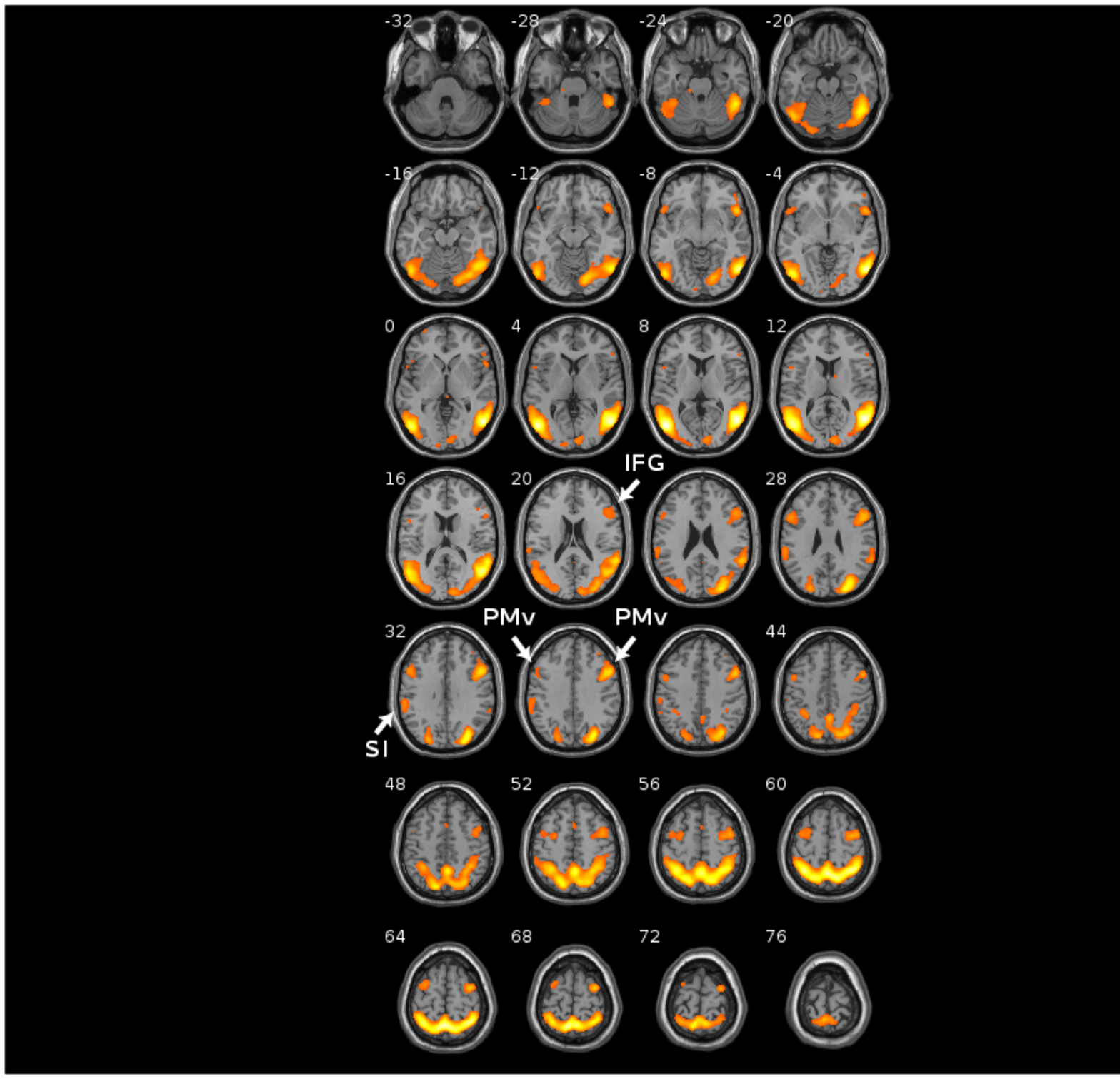}
\endgroup
\caption{Map $\mathbf{a}_{\rm est}$ computed for condition (i) and ``common'' subspace dimension equal to $30$. The map was thresholded such that the 10\% of the voxels with the largest voxel score of $\mathbf{a}_{\rm est}$ are shown. A standard Z-transform is not meaningful, since  $\mathbf{a}_{\rm est}$ satisfies nonnegativity constraints.}
\label{fig:2qa}
\end{figure}

\begin{figure}[t]
\center
\begingroup
\sbox0{\includegraphics{example-image}}
\makeatletter
\Gscale@div\myscale{1.56cm}{\wd0}
\includegraphics[clip,trim={0.05\wd0 0.05\wd0 0.05\wd0 0.00\wd0},scale=\myscale]{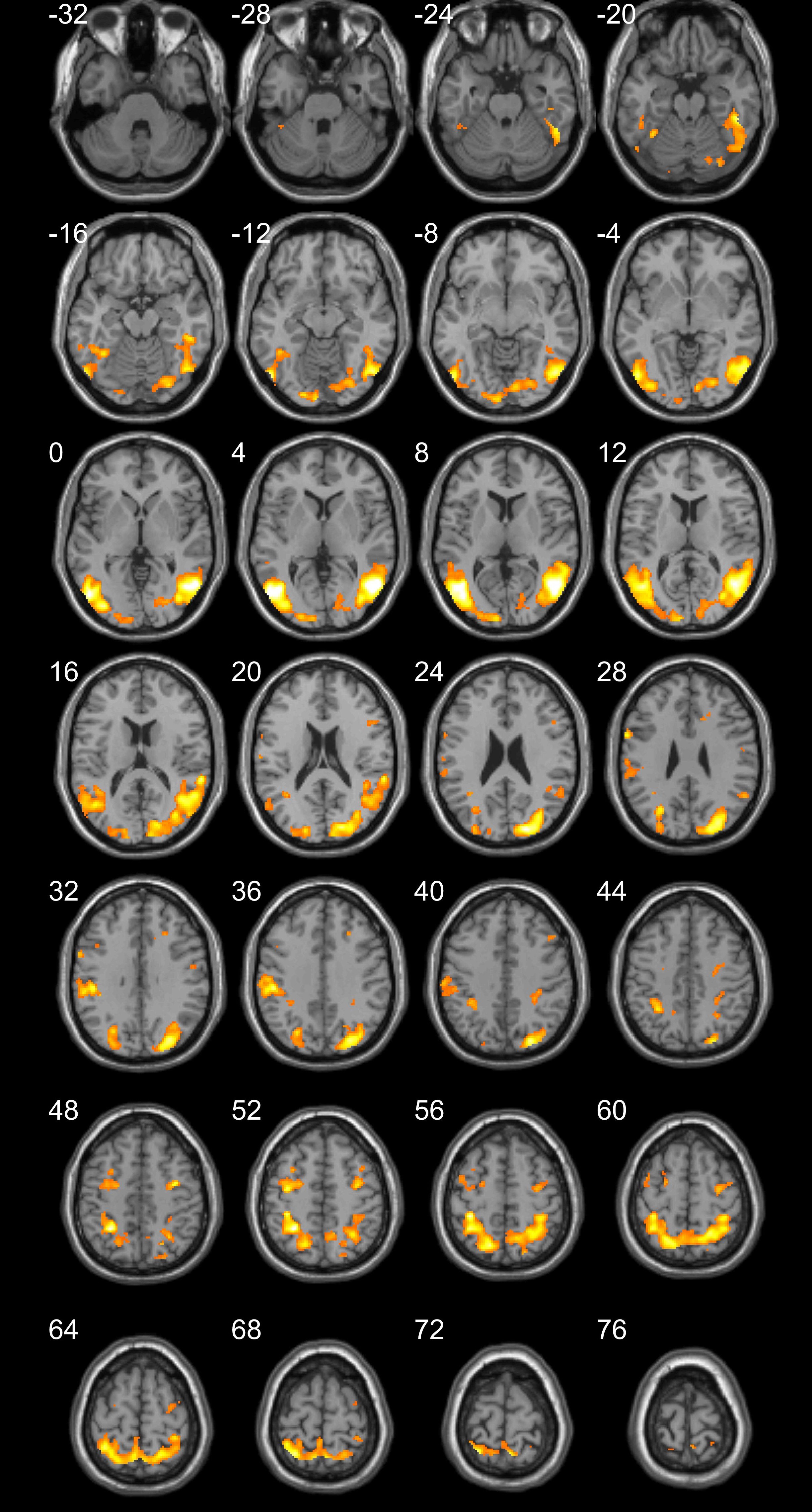}
\endgroup
\caption{Map obtained, for condition (i), using the conventional General Linear Model to the original fMRI data matrices $\left\{\mathbf{X}_k\right\}_{k=1}^K$ with a priori knowledge of the timing of the experimental (video clip observation) and reference blocks (static hand viewing) in SPM (at a standard threshold of $p < 0.001$ uncorrected in the 2nd level analysis).} 
\label{fig:2oa}
\end{figure}

\begin{figure}[t]
\vspace{-0.2mm}
\center
\begingroup
\sbox0{\includegraphics{example-image}}%
\makeatletter
\Gscale@div\myscale{10.81cm}{\wd0}
\includegraphics[clip,trim={0.975\wd0 0.20\wd0 0.68\wd0 0.18\wd0},scale=\myscale]{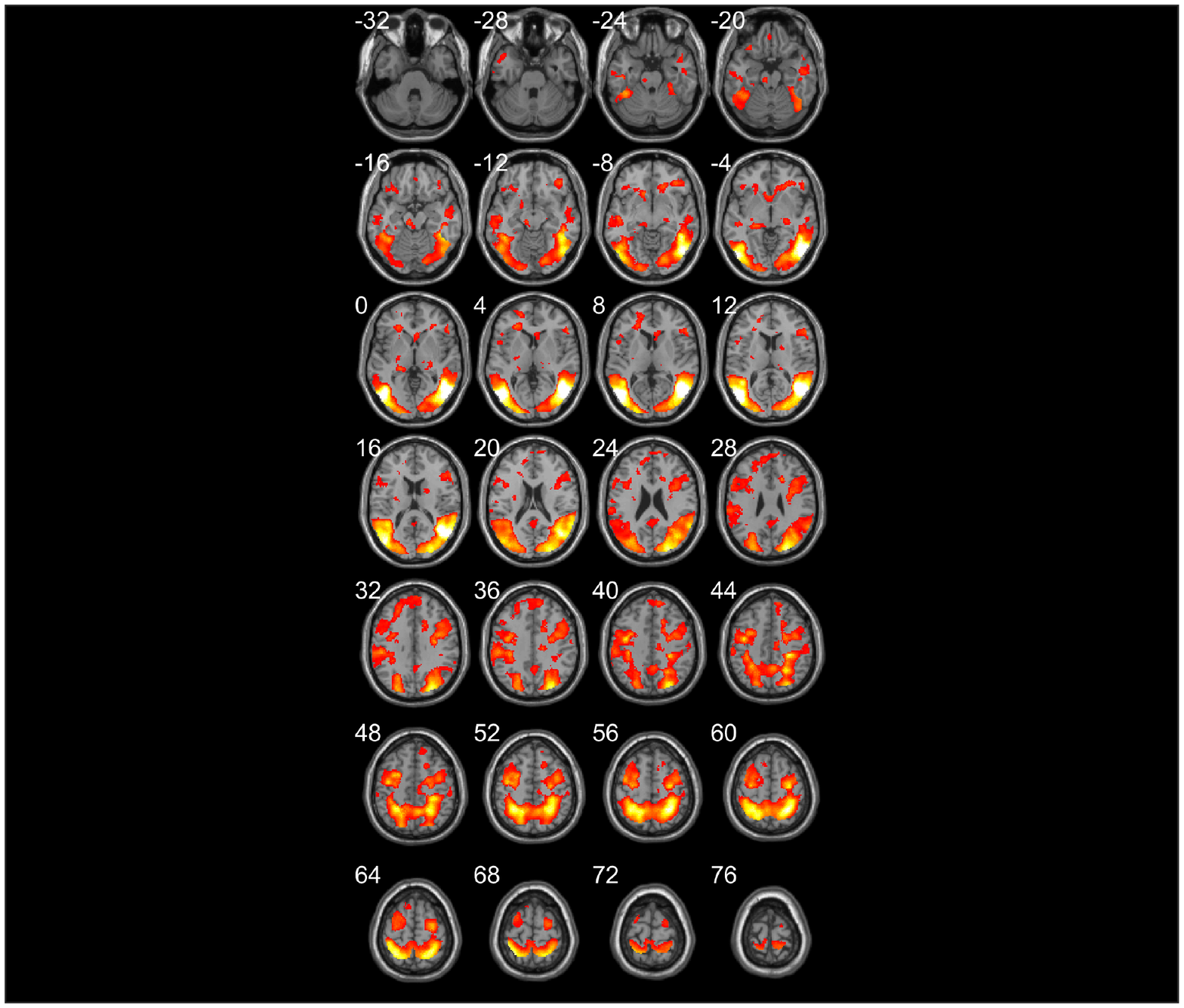}
\endgroup
\caption{Map obtained, for condition (i), using the conventional General Linear Model to the denoised fMRI data matrices $\left\{\mathbf{X}^o_k\right\}_{k=1}^K$ with a priori knowledge of the timing of the experimental (video clip observation) and reference blocks (static hand viewing) in SPM (at a standard threshold of $p < 0.001$ uncorrected in the 2nd level analysis).} 
\label{fig:2o_dea}
\end{figure}

\subsubsection{Results}

Next, we present the results that we obtained from the analysis of the dataset from condition (i). As mentioned before, the results from conditions (ii), (iii), and (iv) are presented in the Supplementary Material. 

In Fig. \ref{fig:1a}, we plot the estimated common temporal component $\mathbf{s}_{\rm est}$ for various values of the common spatial subspace dimension, $R$, as well as the normalized, to unit $2$-norm, expected response $\mathbf{s}_{\rm exp}$, for condition (i). A more direct comparison between the normalized expected response with each one of the estimated common temporal components can be made via Fig. $\ref{fig:2a}$, where we plot the absolute correlation coefficients between $\mathbf{s}_{\rm exp}$ and the estimated common temporal component $\mathbf{s}_{\rm est}$, that emerged for all possible values of $R$.

We observe that, for different values of $R$, the estimated common temporal components, ${\bf s}_{\rm est}$, are very much alike. This implies that our estimate is {\em not}\/ sensitive to the true common subspace dimension, which is unknown, in general. Thus, we can get useful results over a wide range of values of $R$. 
Moreover, the estimated common temporal components, ${\bf s}_{\rm est}$, are quite similar to the
expected signal, ${\bf s}_{\rm exp}$, since their correlation coefficient takes values at about $0.8$ and even higher in some cases. 
We conclude that our method effectively estimates the common temporal component, without any prior knowledge about its shape. 

In order to test the usefulness of the gap function towards the estimation of the common spatial subspace dimension, $R$, we randomly partition the subjects into two sets $\mathcal{K}_1$ and $\mathcal{K}_2$ and compute the gap of the subspaces $\mathcal{S}_1$ and $\mathcal{S}_2$ spanned from the bases computed by the gCCA for assumed dimension $\hat{R}=1,\ldots,73$,\footnote{The length of each voxel's time-series is $75$ and the maximum rank of each ${\bf X}_k$ after de-drifting is $73$.} as described in Subsection 
\ref{Subsection_On_Dimension_of_Common_Subspaces}. In Fig. \ref{fig:2lpa}, we plot the gap function, 
$\rho_{g,2}\left(\mathcal{S}_1,\mathcal{S}_2\right)$, as a function of $\hat{R}$. We observe that the value of the gap becomes practically equal to $1$ for $\hat{R}\succsim 26$, which (1) implies that a low common spatial subspace dimension model is appropriate and (2) provides an estimate for the dimension $R$.

In order to test the validity of assumption (\ref{assumption_empty_intersection}), we again partition the considered dataset, subject-wise, into two random subsets, $\mathcal{K}_1$ and $\mathcal{K}_2$, and 
proceed as follows. First, we solve problem (\ref{common_spatial_gCCA}) for $R=30$ twice, once for each subset, leading to ${\bf G}_i^o$
and $\left\{\mathbf{Q}_k^{o^i}\right\}_{k\in\mathcal{K}_i}$, for $i=1,2$ (notice that Fig. 7 indicates that $R=30$ is an appropriate choice for the considered dataset). Based on $\left\{\mathbf{Q}_k^{o^1}\right\}_{k\in\mathcal{K}_1}$ and  $\left\{\mathbf{Q}_k^{o^2}\right\}_{k\in\mathcal{K}_2}$, we solve two problems analogous to (\ref{MAXVAR}), for common time subspace dimensions $\hat{r}=1,\ldots,30$, i.e. all the possible nontrivial dimensions of the common time subspace. We denote the estimates of the common time subspaces that emerged from the data in $\mathcal{K}_1$ and $\mathcal{K}_2$, as $\mathcal{T}_1$ and $\mathcal{T}_2$, respectively. In Fig. \ref{fig:2ka}, we plot the resulting gap, $\rho_{g,2}\left(\mathcal{T}_1,\mathcal{T}_2\right)$, as a function of the assumed dimension $\hat{r}$. We observe that, for $\hat{r} > 1$, the gap is practically equal to $1$, indicating that the common temporal subspace dimension is equal to one, validating our assumption (\ref{common_time_component}) and, thus, (\ref{assumption_empty_intersection}). 

In Fig. \ref{fig:2lqa}, we plot the intensities across subjects, $\boldsymbol{\lambda}_{\rm est}$, that resulted from the solution of (\ref{Prob_2}), for $R=30$. We observe that the task-related common rank-one term appears in all but one subject.

In Fig. \ref{fig:2qa}, we depict the thresholded spatial map ${\bf a}_{\rm est}$ that emerged from the solution of (\ref{Prob_2}) (only the top $10\%$ of the values have been included). In Fig. \ref{fig:2oa}, we depict the contrast map obtained by applying the conventional General Linear Model (GLM) with a priori knowledge of the timing of the experimental (video clip observation) and reference blocks (static hand viewing) in SPM (at a standard threshold of $p < 0.001$ uncorrected) to the original data. 

We observe that, contrary to the GLM based analysis of the original data, our method successfully captures all clusters of activation voxels in key components of the brain network putatively involved in evaluating the kinematic characteristics and intentions of the observed actions of other subjects, including the inferior frontal gyrus (IFG), ventral Premotor Area (PMv), and primary somatosensory area (SI). 

We claim that the main reason for this phenomenon is the combination of the successful denoising achieved by 
(i) projecting the original data matrices $\left\{\mathbf{X}_k\right\}_{k=1}^K$ onto the common spatial subspace spanned by $\mathbf{G}^o$ and (ii) computing a high quality estimate of the task-related temporal component ${\bf s}_{\rm est}$ via solving problem (\ref{MAXVAR}). 

In order to support this claim, in Fig. \ref{fig:2o_dea}, we plot the spatial map that emerged after applying the conventional General Linear Model with a priori knowledge of the timing of the stimulus in SPM (at the same standard threshold of $p < 0.001$ uncorrected set in creating the corresponding activation map from the original data shown in Fig. \ref{fig:2oa}) to the \textit{denoised} data matrices $\left\{\mathbf{X}_k^o\right\}_{k=1}^K$. We observe that the denoised data significantly increase the sensitivity of the GLM in detecting activations in several regions that are putative key components of the frontoparietal network supporting mental simulation. 

\begin{table*}
\begin{center}
\begin{tabular}{ |P{2cm}||P{3.5cm}|P{3.5cm}|P{3.5cm}|P{2cm}| }
 \hline
 & GLM Original $\cap$ Proposed method & GLM Original $\cap$ GLM Denoised & GLM Denoised $\cap$ Proposed method & Common to all \\
 \hline
 \hline
 condition (i)   & 72.79\% & 78.18\% & 83.81\% & 68.92\%\\
 condition (ii)  & 76.11\% & 81.36\% & 84.68\% & 72.19\%\\
 condition (iii) & 79.91\% & 81.43\% & 94.02\% & 77.98\%\\
 condition (iv)  & 81.03\% & 83.53\% & 91.68\% & 78.83\%\\
 \hline
\end{tabular}
\caption{ Quantitative comparison in terms of spatial overlap between the proposed method and the GLM, after applying it to the original and the denoised data. For GLM, the average beta maps were computed across all subjects and the resulting maps were restricted to the voxels attaining the 10\% of the largest values. For the proposed method, the considered map emerged after restricting the proposed estimate $\mathbf{a}_{\rm est}$ only to the 10\% of its largest values.}
\label{tab:label}
\end{center}
\end{table*}

Finally, in order to compare, in a quantitative manner, the proposed method and GLM we consider the following setup. We apply the GLM with the expected temporal response ($\mathbf{s}_{\rm exp}$) to the original and the denoised data, as defined in relation (21). We compute the average beta maps across all subjects for both datasets and keep the 10\% of the voxels that attained the largest average beta scores. In Table \ref{tab:label}, we present the percentages of pairwise spatial overlaps between the two restricted average beta maps described above and the map that emerges after considering only the $10\%$ of the voxels that attained the largest values of the proposed estimate, $\mathbf{a}_{\rm est}$, for {\em all}\/ the considered real-world datasets (conditions (i)-(iv)). Moreover, in the last column of Table \ref{tab:label}, we present the percentages of the spatial overlaps between all three of the considered maps. 

Based on Table \ref{tab:label}, we can observe that, for all conditions, the following hold:
\begin{itemize}
\item[(i)] 
all the restricted maps have a spatial overlap of approximately 74\%,  
\item[(ii)] 
denoising leads to a difference of approximately 20\% between the GLM-based restricted maps,
\item[(iii)] 
there is a significant difference between the vanilla GLM-based map (original data) and the map resulting from the proposed method,
\item[(iv)] 
the difference mentioned in (iii) becomes consistently smaller when we compare the maps derived from the denoised data with GLM and the proposed method. 
\end{itemize}

\section{Conclusion}

We considered the problem of multi-subject task-related fMRI analysis with one stimulus. We proposed a data generating model which takes into account
both task-related and resting-state common spatial components. We used two successive gCCAs and computed an estimate of the common temporal component, which was then used for the construction of the map of the activated brain regions. We used synthetic data and observed that our estimates are very accurate even at very low SNR. We applied our method to real-world datasets and compared the results with those of a standard GLM procedure. 

We observed that the denoised data (after the projection onto the common spatial subspace) 
lead to improved GLM results, thus, supporting our data generation model and its denoising properties.

\appendices

\section{Compression}
\label{appendix:Compression}

Let $\mathbf{X}_k\in\mathbb{R}^{N\times M}$, for $k=1,\dots,K$, be a set of full column rank matrices. As we showed in section \ref{subsection_common_spatial_CCA}, in order to solve problem (\ref{common_spatial_gCCA}), we have to compute and decompose a $(N\times N)$ matrix, which may be very demanding for large values of $N$ that could emerge in a whole-brain analysis setting. Inspired by the compression technique presented in \cite{stegeman2007comparing}, in this section we show that, if $KM\ll N$, then there is a way to circumvent this problem. 

Let $\mathbf{Y}\in\mathbb{R}^{N\times KM}$ be defined as
\begin{equation}
\mathbf{Y}=\left[\mathbf{X}_1\cdots\mathbf{X}_K\right],
\end{equation}
and consider a factorization of $\mathbf{Y}$ in the form
\begin{equation}
\mathbf{Y}=\mathbf{U}_Y\mathbf{V}_Y,
\end{equation}
such that $\mathbf{U}_Y\in\mathbb{R}^{N\times KM}$ is a columnwise orthonormal matrix, i.e. $\mathbf{U}_Y^T\mathbf{U}_Y=\mathbf{I}_{KM}$, and $\mathbf{V}_Y\in\mathbb{R}^{KM\times KM}$. Then, it holds true that
\begin{equation}
\spn\left(\mathbf{X}_k\right)\subseteq\spn\left(\mathbf{Y}\right)=\bigcup_{k=1}^K\spn\left(\mathbf{X}_k\right).
\end{equation}
Furthermore, we have that
\begin{equation}
\spn\left(\mathbf{Y}\right)\subseteq\spn\left(\mathbf{U}_Y\right).
\end{equation}
As a result, we conclude that, for each matrix $\mathbf{X}_k$, there exists a matrix $\mathbf{H}_k\in\mathbb{R}^{KM\times M}$ such that
\begin{equation}
\mathbf{X}_k = \mathbf{U}_Y\mathbf{H}_k,
\label{Y_k_fac}
\end{equation}
for $k=1,\ldots,K$ and, since $\mathbf{U}_Y^T\mathbf{U}_Y=\mathbf{I}_{KM}$, we also have
\begin{equation}
\mathbf{H}_k = \mathbf{U}_Y^T\mathbf{X}_k,
\end{equation}
for $k=1,\ldots,K$.

Now, recall that matrix $\mathbf{M}$ is defined as 
\begin{equation}
\mathbf{M}=\sum_{k=1}^K\mathbf{X}_k\mathbf{X}_k^{\dagger}=\sum_{k=1}^K\mathbf{X}_k\left(\mathbf{X}_k^T\mathbf{X}_k\right)^{-1}\mathbf{X}_k^T.
\end{equation}
Using $\left(\ref{Y_k_fac}\right)$, we obtain
\begin{equation}
\small
\begin{split}
\mathbf{M}&=\sum_{k=1}^K\mathbf{X}_k\left(\mathbf{X}_k^T\mathbf{X}_k\right)^{-1}\mathbf{X}_k^T\\
&=\mathbf{U}_Y\left(\sum_{k=1}^K\mathbf{H}_k\left(\mathbf{H}_k^T\mathbf{H}_k\right)^{-1}\mathbf{H}^T_k\right)\mathbf{U}_Y^T\\
&=\mathbf{U}_Y\tilde{\mathbf{M}}\mathbf{U}_Y^T,
\end{split}
\end{equation}
where 
\begin{equation}
\tilde{\mathbf{M}}:=\sum_{k=1}^K\mathbf{H}_k\left(\mathbf{H}_k^T\mathbf{H}_k\right)^{-1}\mathbf{H}^T_k\in\mathbb{R}^{KM\times KM}.
\label{tilde_M_def}
\end{equation}
Let the eigenvalue decompositions of $\mathbf{M}$ and $\tilde{\mathbf{M}}$ be
\begin{equation}
\mathbf{M}=\mathbf{U}_{M}\boldsymbol{\Lambda}_{M}\mathbf{U}_{M}^T,~~ \tilde{\mathbf{M}}={\mathbf{U}}_{\tilde{M}}\boldsymbol{\Lambda}_{\tilde{M}}\mathbf{U}_{\tilde{M}}^T.
\end{equation}
Since $\mathbf{M}=\mathbf{U}_Y\tilde{\mathbf{M}}\mathbf{U}_Y^T$ and $\mathbf{U}_Y^T\mathbf{U}_Y=\mathbf{I}_{KM}$, we have that
\begin{equation}
\boldsymbol{\Lambda}_{M}=\boldsymbol{\Lambda}_{\tilde{M}},\quad\mathbf{U}_{M} = \mathbf{U}_Y\mathbf{U}_{\tilde{M}}.
\end{equation}
Recall that we are interested in the eigenvectors associated with the $R$ largest eigenvalues of $\mathbf{M}$, since
${\bf G}^o=\mathbf{U}_{M}(:,1:R)$.
Furthermore, we have that
\begin{equation}
\mathbf{U}_{M}\left(:,1:R\right)=\mathbf{U}_Y\mathbf{U}_{\tilde{M}}\left(:,1:R\right).
\label{U_M_from_U_Y_Y_Mt}
\end{equation}
Thus, it suffices to solve the MAXVAR problem for the ``low dimensional'' matrices $\mathbf{H}_k$ and compute $\mathbf{U}_{\tilde{M}}\left(:,1:R\right)$. Then, we use (\ref{U_M_from_U_Y_Y_Mt}) and obtain $\mathbf{G}^o$ without the direct 
computation of matrix $\mathbf{M}$.

\section*{Acknowledgment}
{ The authors would like to thank Dr. Eleftherios Kavroulakis for helping with data preprocessing and Prof. Michalis Zervakis for useful comments.}

\ifCLASSOPTIONcaptionsoff
  \newpage
\fi

\bibliographystyle{IEEEtran}
\bibliography{Multisubject_task-related_fMRI_data_processing_via_a_two-stage_generalized_canonical_correlation_analysis_arxiv}

\end{document}